\documentclass[%
 reprint,
superscriptaddress,
 amsmath,amssymb,
 aps,
pra,
floatfix,
]{revtex4-2}

\usepackage{graphicx}
\usepackage{dcolumn}
\usepackage{bm}

\usepackage{amsmath,amssymb,amsfonts}
\usepackage{xcolor}

\usepackage{hyperref}
\usepackage{braket}

\usepackage[normalem]{ulem}
\newcommand\erase{\bgroup\markoverwith{\textcolor{red}{\rule[.5ex]{2pt}{0.4pt}}}\ULon}
\newcommand\eraseblue{\bgroup\markoverwith{\textcolor{blue}{\rule[.5ex]{2pt}{0.8pt}}}\ULon}

\begin{document}


\title{A Study on Quantum Car-Parrinello Molecular Dynamics with Classical Shadows\\ for Resource Efficient Molecular Simulation}

\author{Honomi Kashihara}\thanks{These authors share first authorship.}
\affiliation{Department of Mechanical Engineering, Keio University, Hiyoshi 3‑14‑1, Kohoku, Yokohama 223‑8522, Japan}
\author{Yudai Suzuki}\thanks{These authors share first authorship.}
\affiliation{Quantum Computing Center, Keio University, Hiyoshi 3‑14‑1, Kohoku,Yokohama 223‑8522, Japan}
\author{Kenji Yasuoka}
\affiliation{Department of Mechanical Engineering, Keio University, Hiyoshi 3‑14‑1, Kohoku, Yokohama 223‑8522, Japan}

\begin{abstract}
\textit{Ab-initio} molecular dynamics (AIMD) is a powerful tool to simulate physical movements of molecules for investigating properties of materials.
While AIMD is successful in some applications, circumventing its high computational costs is imperative to perform large-scale and long-time simulations.
In recent days, near-term quantum computers have attracted much attentions as a possible solution to alleviate the challenge.
Specifically, Kuroiwa et al. proposed a new AIMD method called quantum Car-Parrinello molecular dynamics (QCPMD), which exploits the Car-Parrinello method and Langevin formulation to realize cost-efficient simulations at the equilibrium state, using near-term quantum devices.
In this work, we build on the proposed QCPMD method and introduce the classical shadow technique to further improve resource efficiency of the simulations.
More precisely, classical shadows are used to estimate the forces of all nuclei simultaneously, implying this approach is more effective as the number of molecules increases. 
We numerically study the performance of our scheme on the $\text{H}_2$ molecule and show that QCPMD with classical shadows can simulate the equilibrium state.
Our results will give some insights into efficient AIMD simulations on currently-available quantum computers.

\end{abstract}
%
\maketitle
%
%
\section{Introduction}
\textit{Ab-initio} molecular dynamics (AIMD) is an effective computational tool to elucidate properties of molecules by examining their behavior at the atomic level~\cite{tuckerman2002ab,iftimie2005ab,marx2009ab,paquet2018computational}. 
With the AIMD method, a number of studies have attempted to clarify the underlying dynamics of material properties such as chemical reactions~\cite{mosconi2015ab}, diffusion~\cite{yang2011li,he2018statistical}, amorphous materials~\cite{johari2011mixing} and vibrational frequencies~\cite{thomas2013computing}.
On the other hand, AIMD requires quantum-mechanical computations of potential energy surfaces, which is computationally demanding and thus hiders large-scale and long-time simulations.
Hence, seeking new approaches to resolve the computational cost problem is imperative.

A possible solution to circumvent the difficulty is quantum computing, which can potentially outperform classical computers in terms of the information processing speed. 
Unfortunately, ideal fault-tolerant quantum computers are not available now; that is, we cannot execute quantum algorithms with theoretical guarantees such as Ref.~\cite{shor1994algorithms,kitaev1995quantum,grover1996fast,harrow2009quantum} on currently-available quantum computers for practical applications.
To be more specific, near-term quantum computers, the so-called noisy intermediate-scale quantum (NISQ) devices, are limited in the number of qubits and suffer from noise through computation~\cite{preskill2018quantum}.
This suggests the challenge of NISQ computers even for simple computations.
Nevertheless, recent works spotlight the potential power of the NISQ devices.
An example includes experimental demonstrations of a sampling task where NISQ devices can be superior to classical means~\cite{arute2019quantum}.
Fueled by such results, many attempts have been made to explore the utility of NISQ devices by e.g., utilizing hybrid quantum-classical strategies~\cite{cerezo2021variational}.

In the literature of AIMD, the power of NISQ devices has also been explored.
Originally proposed is a hybrid quantum-classical method~\cite{fedorov2021ab} where the electronic ground state is computed by variational quantum eigensolvers (VQEs)~\cite{peruzzo2014variational} on NISQ devices, whereas the update of nuclei positions is executed on classical computers.
We note that the VQE algorithm is performed using parameterized quantum circuits (PQCs) whose parameters are also tuned by classical optimizers.
This method was proposed based on the expectation that VQE can solve the ground state problem more efficiently than classical methods.
On the other hand, there is room for investigation in the practicality of the method. 
For example, this proposal does not take into account the inevitable statistical noise caused by a finite number of measurement shots. 
In addition, performing VQE algorithms with sufficient accuracy at each iteration demands many rounds of quantum state preparation and measurements.
To tackle these challenges, Ref.~\cite{kuroiwa2022quantum} recently proposed a new AIMD method called Quantum Car-Parrinello molecular dynamics (QCPMD).
The key contributions of QCPMD are two-fold: (1) inspired by the idea of classical Car-Parrinello molecular dynamics, the subroutine of VQE is replaced with parallel updates of nuclei positions and parameters in PQCs, and (2) statistical noise is rather utilized as thermostats with the help of Langevin dynamics formula.
Thanks to these modifications, QCPMD can perform cost-efficient simulation of molecules at equilibrium using NISQ devices, while allowing for inaccurate calculations of the electronic ground state and the statistical noise by measurement.

In this work, we build on the QCPMD method proposed in Ref.~\cite{kuroiwa2022quantum} and introduce classical shadows to improve its resource efficiency. 
The classical shadow technique is a powerful tool to simultaneously estimate expectation values of multiple observables~\cite{huang2020predicting}. 
Hence, our idea is to utilize the technique for reducing the number of samples required to estimate forces on the nuclei at each step.
More specifically, we make use of the classical snapshots of quantum states to obtain $3N$ forces of $N$ nuclei simultaneously at each iteration, instead of naive estimations of the forces.
This indicates that our scheme would be more effective than the naive approach as the number of nuclei increases.
We then consider a $\text{H}_2$ molecule as a testbed and numerically show that QCPMD with classical shadows can successfully simulate the equilibrium state.
These results will push forward the practical use of NISQ devices for efficient AIMD simulation.

The remaining of this paper is structured as follows.
In Section~\ref{sec:method}, we provide a brief overview of QCPMD and classical shadows, followed by explanation of our scheme.
Then, we show numerical results on simulation of a $\text{H}_2$ molecule in Section~\ref{sec:result}.
Lastly, we conclude this work in Section~\ref{sec:conc}.

\section{Methods} \label{sec:method}
In what follows, we show preliminaries on AIMD simulations, QCPMD~\cite{kuroiwa2022quantum} and classical shadows~\cite{huang2020predicting}.
We then elaborate on our framework and its possible advantages and drawbacks.

\subsection{Preliminaries}
\subsubsection{Brief Overview of AIMD Simulations}
In AIMD simulations, the motion of electrons and nuclei are solved separately based on the Born-Oppenheimer approximation~\cite{ANDP:ANDP19273892002}.
This approximation assumes that electronic and nuclear motion can be decoupled because electrons are much lighter and hence moves faster than nuclei.
With this approximation, the time evolution of molecules is made tractable.

More precisely, AIMD treats nuclei as classical point masses, while electrons are still handled quantum-mechanically.
In other words, at each iteration, the Hamilton's canonical equations are solved for the nuclei motion, where the forces in use are obtained from the potential energy surfaces we computed through the solution of the Schr\"{o}dinger equation for the elections.
The quantum-classical approach in AIMD can enable more accurate simulations than the classical molecular dynamics simulation that uses the empirical and simplified potential functions called force field for calculating the potential energy surface.
On the other hand, solving the Schr\"{o}dinger equation at each step is computationally demanding and thus AIMD is challenging for long-time simulations with many molecules.
Hence, attentions have been paid to methods for circumventing the difficulty.

Car and Parrinello proposed an approach that mitigates the cost problem for computing eigenstates of the electrons~\cite{car1985unified}.
In a broad sense, the key feature of the Car-Parrinello molecular dynamics is to treat the electronic wavefunctions and nuclear positions as dynamical variables and then evolve them together at each timestep.
While the success of Car-Parrinello molecular dynamics requires the adiabaticity condition (i.e., a condition that electronic state can instantaneously follow the change of nuclei positions), avoiding the eigenvalue problem can largely enhance the efficiency of molecular simulations.

\subsubsection{Quantum Car-Parrinello Molecular dynamics}
In the following, we review Quantum Car-Parrinello molecular dynamics (QCPMD), an AIMD method that exploits NISQ computers.
Actually, the QCPMD is not the first attempt to utilize NISQ devices for AIMD simulations.
In the literature, Ref.~\cite{fedorov2021ab,sokolov2021microcanonical} straightforwardly employed the VQE algorithms for AIMD simulations to compute the electronic ground state.
We remind that VQE solves eigenvalue problems and is amenable to implementation on NISQ devices~\cite{peruzzo2014variational}; then, the hope is that VQE is more efficient than conventional approaches.
The AIMD with VQE (VQE-AIMD) is highly related to the QCPMD and thus we first give its overview briefly.

In the VQE-AIMD, we prepare a quantum state $\ket{\Psi(\bm{\theta})}=U(\bm{\theta})\ket{\Psi_{0}}$ generated by applying a parameterized quantum circuit (PQC) $U(\bm{\theta})$ to an initial state $\ket{\Psi_{0}}$.
Then, we optimize the parameters $\bm{\theta}$ so that the parameterized quantum state can approximate the ground state of a Hamiltonian $H(\bm{R})$ of target $N$ molecules.
Here, $\bm{R}=(\bm{R}_1,\ldots,\bm{R}_{N})^{\mathsf{T}}$ denotes the nuclei positions with the Cartesian coordinates of $l$-th nucleus $\bm{R}_{l} = (R_{l,x},R_{l,y},R_{l,z})^{\mathsf{T}}$.
A common strategy to obtain the ground state is minimize the energy 
\begin{equation} \label{eq:pes}
    L(\bm{R},\bm{\theta}) = \mathrm{Tr}\left[ H(\bm{R})\ket{\Psi(\bm{\theta})}\bra{\Psi(\bm{\theta})}\right]
\end{equation}
as small as possible according to the variational principle.
Once one can obtain the ground state energy, the forces on nuclei are also computable: the force $F_{l,\alpha}(\bm{R},\bm{\theta})$ on the $l$-th nucleus in the $\alpha$ direction is expressed as
\begin{equation} \label{eq:f_r}
\begin{split}
    F_{l,\alpha}(\bm{R},\bm{\theta}) &= - \frac{\partial}{\partial R_{l,\alpha}} L(\bm{R},\bm{\theta})\\
    &\approx - \mathrm{Tr}\left[ \frac{d H(\bm{R})}{d_{l,\alpha}}\ket{\Psi(\bm{\theta})}\bra{\Psi(\bm{\theta})}\right],
\end{split} 
\end{equation}
where we consider the Hellman-Feynmann force only and ignore the Puley force (i.e., $\mathrm{Tr}[ H(\bm{R}) \frac{\partial}{\partial R_{l,\alpha}} (\ket{\Psi(\bm{\theta})}\bra{\Psi(\bm{\theta})}) ]$)~\cite{pulay1969ab}, assuming the ground state is well-approximated.
Note that we obtain the Hellman-Feynmann force by estimating the expectation values of the corresponding observables, the gradient of the Hamiltonian $d H(\bm{R})/d_{l,\alpha}$.
Also, the force shown in the right hand side of Eq.~\eqref{eq:f_r} can be obtained by taking the finite difference
\begin{equation}
\begin{split}
    &\mathrm{Tr}\left[ \frac{d H(\bm{R})}{d_{l,\alpha}}\ket{\Psi(\bm{\theta})}\bra{\Psi(\bm{\theta})}\right] \\
    &\approx \frac{\mathrm{Tr}\left[ \left(H(\bm{R}+d \bm{e}_{l,\alpha}) - H(\bm{R}-d \bm{e}_{l,\alpha})\right)\ket{\Psi(\bm{\theta})}\bra{\Psi(\bm{\theta})}\right]}{2d}
\end{split}
\end{equation}
with the unit vector $\bm{e}_{l,\alpha}$ and a scalar value $d$.
Using the forces $\bm{F}(\bm{R},\bm{\theta})=(\bm{F}(\bm{R}_1,\bm{\theta}),\ldots, \bm{F}(\bm{R}_N,\bm{\theta}))^{\mathsf{T}}$ with $\bm{F}(\bm{R}_l,\bm{\theta})=(F_{l,x}(\bm{R},\bm{\theta}),F_{l,y}(\bm{R},\bm{\theta}),F_{l,z}(\bm{R},\bm{\theta}))^{\mathsf{T}}$ for fixed $\bm{\theta}$, we solve the classical equations of motions
\begin{equation} \label{eq:EOM}
    \bm{m} \ddot{\bm{R}} = \bm{F} (\bm{R},\bm{\theta})
\end{equation}
with mass $\bm{m}$ and then update the nuclei positions.
This procedure is repeated until the terminating condition is met.

The QCPMD is in spirit similar to the VQE-AIMD, but its efficiency in samples is improved by leveraging the idea of Car-Parrinello molecular dynamics and Langevin dynamics formulation.
The first difference is that the VQE is not employed in QCPMD.
That is, similar to the classical Car-Parrinello molecular dynamics, the parameters $\bm{\theta}$ in the PQC are also treated as dynamical variables.
Then, the nuclei configurations $\bm{R}$ and the parameters $\bm{\theta}$ are separately updated in parallel.
This modification is adopted to avoid a problem in VQE that a large number of optimization rounds (equivalently, state preparations and measurements) may be needed to obtain the optimal parameters $\bm{\theta}^{*}$.
Secondly, QCPMD takes into account the effect of statistical noise caused by a finite number of measurements.
In practical settings, it is impossible to ignore the statistical noise for estimating observables such as the ground state energy on quantum computers.
Thus, Langevin formulation is adopted to derive a simulation method that rather makes use of the statistical noise as thermostats; namely, the effect of noise is incorporated into the formulation in such a way that it can be used for controlling the temperature of the system.
Thanks to this, we may be able to reduce the sampling cost compared to the VQE-AIMD.
We omit the derivation in this manuscript, but the update rules that take into account the above-mentioned modifications are expressed as follows (please refer to Ref.~\cite{kuroiwa2022quantum} for more details);
\begin{equation} \label{eq:r}
    \bm{R}^{(k)} = \bm{R}^{(k-1)} + \bm{v}^{(k-1)} \Delta t,
\end{equation}
\begin{equation} \label{eq:v}
\begin{split}
    \bm{v}^{(k)} &= (1-\bm{\gamma}(\bm{R}^{(k-1)},\bm{\theta}^{(k-1)}) \Delta t) \bm{v}^{(k-1)}\\
    & \qquad +  \frac{\bm{F} (\bm{R}^{(k-1)},\bm{\theta}^{(k-1)})}{\bm{m}} \Delta t,
\end{split}
\end{equation}
\begin{equation} \label{eq:theta}
    \bm{\theta}^{(k)} = \bm{\theta}^{(k-1)} + \bm{\xi}^{(k-1)} \Delta t,
\end{equation}
\begin{equation} \label{eq:xi}
\begin{split}
    \bm{\xi}^{(k)} &= (1-\bm{\zeta}(\bm{R}^{(k)},\bm{\theta}^{(k-1)}) \Delta t) \bm{\xi}^{(k-1)}\\
    & \qquad +  \frac{\bm{F}_{\bm{\theta}} (\bm{R}^{(k)},\bm{\theta}^{(k-1)})}{\bm{\mu}} \Delta t,
\end{split}
\end{equation}
where $\bm{R}^{(k)}$, $\bm{\theta}^{(k)}$, $\bm{v}^{(k)}=\dot{\bm{R}}^{(k)}$ and $\bm{\xi}^{(k)}=\dot{\bm{\theta}}^{(k)}$ represent the quantities at $k$-th time step, respectively. 
$\Delta t$ represents the time step of the simulation.
Since we regard the parameters $\bm{\theta}$ as dynamical variables, we newly introduce the virtual mass $\bm{\mu}$ and denote the forces on the parameters $\bm{\theta}$ as $\bm{F}_{\bm{\theta}}=-\nabla_{\bm{\theta}} L(\bm{R},\bm{\theta})$. 
Also, $\bm{\gamma}(\bm{R},\bm{\theta})$ and $\bm{\zeta}(\bm{R},\bm{\theta})$ denote the coefficients determining the strength of dissipation for $\bm{v}$ and $\bm{\xi}$, respectively.
These coefficients are respectively defined as follows;
\begin{equation} \label{eq:gamma}
    \bm{\gamma}(\bm{R},\bm{\theta}) = \frac{\bm{f}^2(\bm{R},\bm{\theta})\beta \Delta t}{2\bm{m}},
\end{equation}
\begin{equation} \label{eq:zeta}
    \bm{\zeta}(\bm{R},\bm{\theta}) = \frac{\bm{f}_{\bm{\theta}}^2(\bm{R},\bm{\theta})\beta \Delta t}{2\bm{\mu}},
\end{equation}
where $\beta=1/k_{b}T$ with Boltzmann constant $k_{B}$ is the inverse temperature and $\bm{f}^2(\bm{R},\bm{\theta})$ ($\bm{f}_{\bm{\theta}}^2(\bm{R},\bm{\theta})$) is the statistical variance of the forces on nuclei (parameters).
We remark that, according to Ref.~\cite{kuroiwa2022quantum}, the statistical noise is absorbed into the coefficients and hence could play a role in the temperature control.
\\

\subsubsection{Classical Shadows}
The classical shadow technique enables us to predict many properties of a target quantum state~\cite{huang2020predicting}. 
More concretely, the order $\log(M)$ size of samples is sufficient for classical shadows to predict $M$ target linear functions.
The main idea of this method is to perform a series of randomized measurements on multiple copies of the quantum state.
In the following, we explain the detailed procedure to reconstruct the underlying quantum state.
Here, our target is a $n$-qubit quantum state represented as $\rho = \ket{\Psi}\bra{\Psi}$ in the density matrix representation.
\begin{enumerate}
    \item Apply a random unitary $U$ to the quantum state $\rho$, i.e., $\rho \to U\rho U^{\dagger} $. 
    Note that the unitary is randomly selected from an ensemble of unitaries $\mathcal{U}$.
    \item Measure all qubits in the computational basis and store the measured bit-string $\ket{\hat{s}}\in\{0,1\}^{n}$ as classical description of $U^{\dagger} \ket{\hat{s}} \bra{\hat{s}} U$.
    \item Apply the inverted quantum channel $\mathcal{M}^{-1}$ to $U^{\dagger} \ket{\hat{s}} \bra{\hat{s}} U$ to produce the classical snapshot $\hat{\rho}=\mathcal{M}^{-1}(U^{\dagger} \ket{\hat{s}} \bra{\hat{s}} U)$ of a target quantum state $\rho$.
    As we give a concrete example later, the quantum channel depends on the unitary ensemble.
    Also, this post-processing is fully classical.
    Then, by construction, we can recover the quantum state $\rho$ by the classical shadow in expectation, i.e., $\rho = \mathbb{E}_{\hat{s}} \mathbb{E}_{U} [\hat{\rho}]$.
    \item Repeat this procedure $N_{S}$ times to get the empirical average $\tilde{\rho}=\frac{1}{M}\sum_{j=1}^{N_{S}} \hat{\rho}^{(j)}$, which results in the exact quantum state $\rho$ in the limit of $M\to\infty$.
\end{enumerate}
We remark that, in case the random Pauli-basis measurement is performed (i.e., $\mathcal{U}=\{H, HS^{\dagger}, I\}^{\otimes n} $ with the identity gate $I$, the Hadamard gate $H$ and the phase gate $S$), the classical snapshot is expressed as 
\begin{equation}
    \hat{\rho} = \bigotimes_{i=1}^{n} (3U_{i}^{\dagger}\ket{\hat{s}_{i}}\bra{\hat{s}_{i}}U_{i}-I)
\end{equation}
with $\hat{s}=\hat{s}_1\hat{s}_2\ldots\hat{s}_n$ and $U=\otimes_{i=1}^{n}U_{i}$.

Moreover, we can estimate expectation values of many observables $\{O_i\}$ with the classical snapshots at hand: that is, $\mathrm{Tr}[O_i\rho]\approx \frac{1}{N_{S}} \sum_{j=1}^{N_{S}} \mathrm{Tr}[O_i\hat{\rho}^{(j)}] $.
We note that we can also employ the median-of-means protocol~\cite{huang2020predicting} for better estimation, i.e., $\mathrm{Tr}[O_i\rho] = \text{median} \{\mathrm{Tr}[O_i\hat{\rho}^{1}],\ldots,\mathrm{Tr}[O_i\hat{\rho}^{K}]\}$ with classical shadows constructed from $K$ snapshots $\hat{\rho}^{k}=\frac{1}{\lfloor N_{S}/K\rfloor } \sum_{j=(k-1) \lfloor N_{S}/K\rfloor +1 }^{k \lfloor N_{S}/K\rfloor} \hat{\rho}^{(j)}$.
Also, the expectation values of multiple observables can be obtained without the explicit construction of the classical snapshots in case the number of qubit $n$ is large.

\subsection{Our Method: QCPMD with Classical Shadows}

Now, we are in a good position to provide the details of our scheme.
In the AIMD simulations, our scheme also uses the update rules of QCPMD in Eqs.~\ref{eq:r} --~\ref{eq:xi}.
On the other hand, we use the classical shadow technique to estimate the forces on the nuclei $\bm{F}(\bm{R},\bm{\theta})$ to improve the resource efficiency.
As stated in the previous section, we need to estimate $3N$ forces on the nuclei of $N$ molecules at each step, indicating the number of quantum state preparations required for a naive approach scales $\mathcal{O}(3NN_{shot})$ with the number of measurement shots for each force $N_{shot}$.
This cost can be further reduced by the classical shadow, because the quantum state used to estimate the forces in Eq.~\eqref{eq:f_r} is the same regardless of the nuclei and its direction.
Namely, we can estimate all the forces, once we get sufficient number of classical snapshots.
Thus, the number of samples required can be reduced to $\mathcal{O}(N_{S})$, which is independent of the number of molecules $N$.
This will improve the resource efficiency of AIMD simulations and harness the applicability of QCPMD for large systems.

Let us note that we can use the classical shadow technique for estimating forces on parameters $\bm{F}_{\bm{\theta}}(\bm{R},\bm{\theta})$, but the improvement would not be expected.
In gradient estimations, a common choice is the parameter shift rule~\cite{mitarai2018quantum,schuld2019evaluating}, where parameter-shifted quantum states are used to compute the gradient, i.e., $\partial L(\bm{R},\bm{\theta})/\partial \theta_i = \frac{1}{2}(L(\bm{R},\bm{\theta}+\frac{\pi}{2}\bm{e}_{i})-L(\bm{R},\bm{\theta}-\frac{\pi}{2}\bm{e}_{i}))$.
This suggests that we need the estimation of $\rho(\bm{\theta}+\frac{\pi}{2}\bm{e}_{i})$ and $\rho(\bm{\theta}-\frac{\pi}{2}\bm{e}_{i})$ for every parameter $\theta_i$.
Therefore, we would not expect the improvement in the resource efficiency with respect to the number of parameters $N_{p}$.

We also remark that, when it comes to predicting expectation values of fixed many Pauli operators, the classical shadow technique with randomized measurement might not gain the advantages in quantum resources for accurate predictions.
This is because some quantum resources could be wasted for certain observables due to the randomness in choosing the basis.
This indicates that deterministic measurements would be more effective when we know what to measure in advance.
From this standpoint, grouping of commuting multi-qubit Pauli~\cite{yen2023deterministic}, the so-called ``derandomized" classical shadows~\cite{huang2021efficient} and  the combination of these two approaches~\cite{gresch2023guaranteed} would be more effective in this setting.
Actually, the settings of QCPMD simulations fall into this situation; we know the observables that we want to estimate and the Hamiltonian $H(\bm{R})$ is given as a weighted sum of Pauli operators $H=\sum_{P\in\{I,X,Y,Z\}^{\otimes n}}w_P P$.
Thus, for accurate prediction, these methods would be useful.
However, we underscore that noise can be absorbed into the temperate control via the Langevin formulation in QCPMD.
This implies that inaccurate estimations by randomized classical shadow can be allowed in the QCPMD simulations, and our scheme is still beneficial because of the reduction in samples from $\mathcal{O}(3NN_{shot})$ to $\mathcal{O}(N_{S})$.\\

\begin{figure}[t]
    \centering
    \includegraphics[scale=0.7]{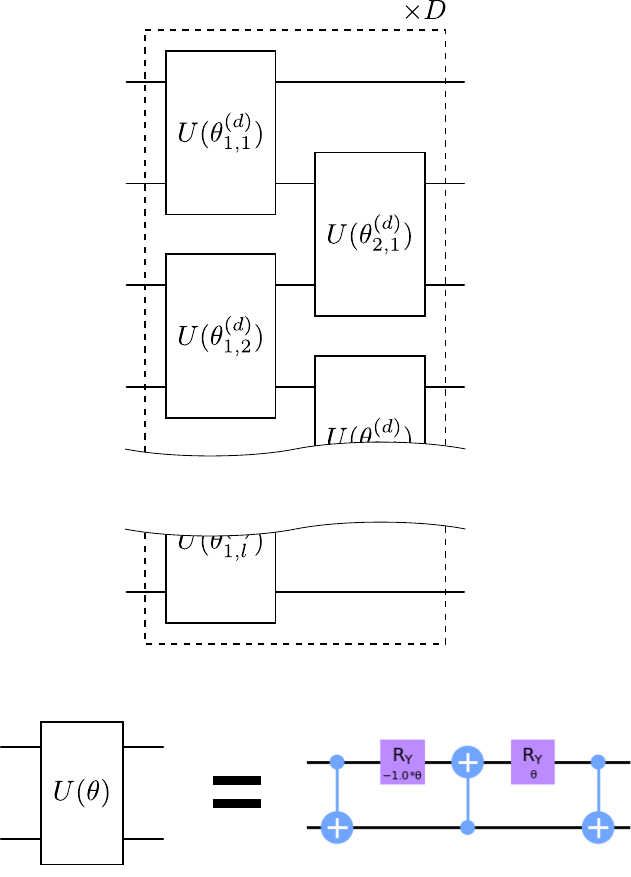}
    \caption{Quantum circuit representation of the RSP type ansatz used in Ref.~\cite{gard2020efficient}. A bunch of gates in the dashed block is repeated $D$ times.}
    \label{fig:rsp}
\end{figure}

\section{Results \& Discussion} \label{sec:result}
Here, we numerically study the performance of QCPMD with and without classical shadows for simulations of a $\text{H}_2$ molecule.
We note that classical shadows are employed to estimate the force of parameters as well for our scheme.
For all the simulation below, Qiskit~\cite{qiskit2024} is used for quantum circuit simulation.
Also, at each step, we use PySCF to prepare fermionic second-quantized Hamiltonians for the electrons with the STO-3G basis set; then, we applied the Jordan-Wigner transformation to convert fermionic Hamiltonians into qubit Hamiltonians.

Our setting is as follows.
We simulate the time evolution of the $\text{H}_2$ molecule at a temperature of $70$ K with the time steps of $\Delta t=0.1$ fs.
The virtual mass for the parameters $\bm{\theta}$ is set to $\mu=0.1$ for calculating the time evolution.
As for the PQC, we use a real-valued symmetry-preserving (RSP) type ansatz~\cite{gard2020efficient} with the depth $D=4$ (Fig.~\ref{fig:rsp}).
We employ the median-of-mean protocol for the QCPMD with classical shadows using $N_{S}=51$ snapshots and $K=3$, while each Pauli operator is measured individually for the ordinary QCPMD with $N_{shot}=51$ measurement shots; sample numbers for QCPMD without classical shadows are larger than the other because every Pauli term for the forces of all coordinates is estimatable with the snapshots for classical shadows.
The coefficients for the dissipation terms are $\gamma=0.8$ and $\zeta=0.8$.
We note that these coefficients depend on the nuclei positions and parameters at that step, and thus should be determined by the equality in Eq.~\eqref{eq:gamma} and Eq.~\eqref{eq:zeta}.
However, we consider the constant value to avoid the computation of variance.
We describe the effect later in this section.

\begin{figure}[t]
    \centering
    \includegraphics[scale=0.48]{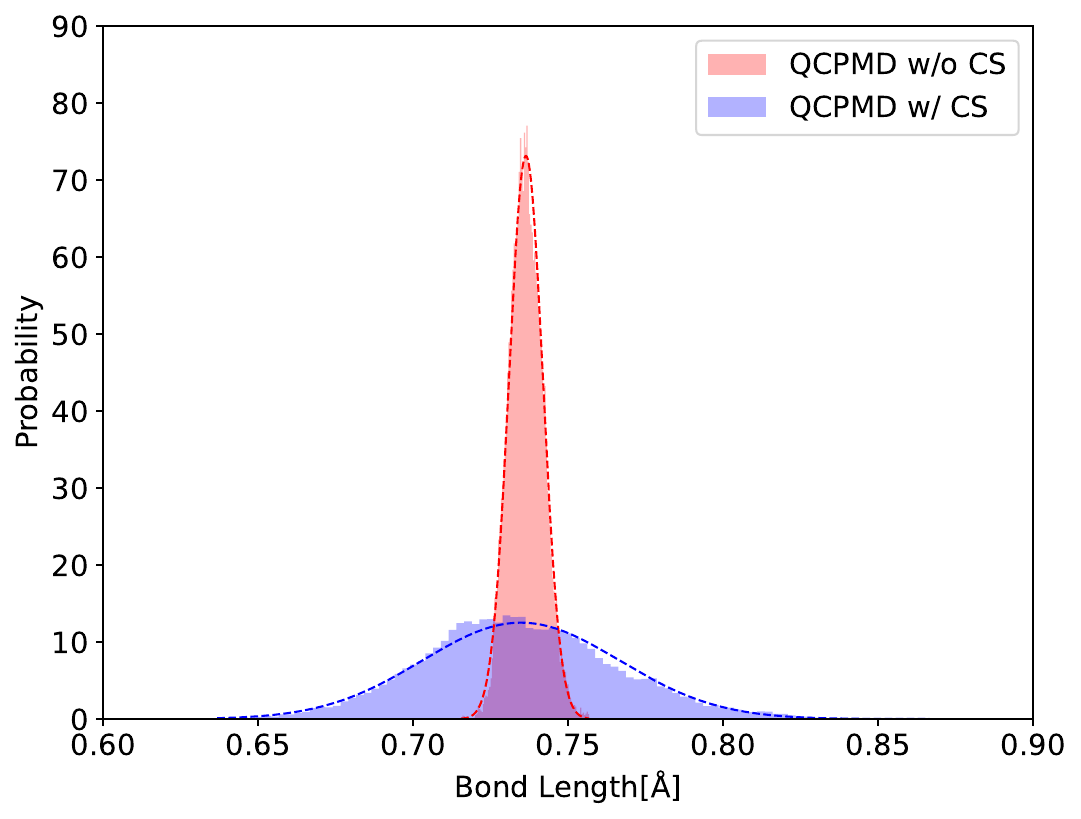}
    \caption{Distributions of bond lengths obtained by the original QCPMD (QCPMD w/o CS) and QCPMD with classical shadows (QCPMD w/ CS) in case the simulations start from the equilibrium state. }
    \label{fig:res1}
\end{figure}

In this study, we focus on two conditions for the initial configuration of the molecule and values of parameters.

\begin{figure}[t]
    \centering
    \includegraphics[scale=0.48]{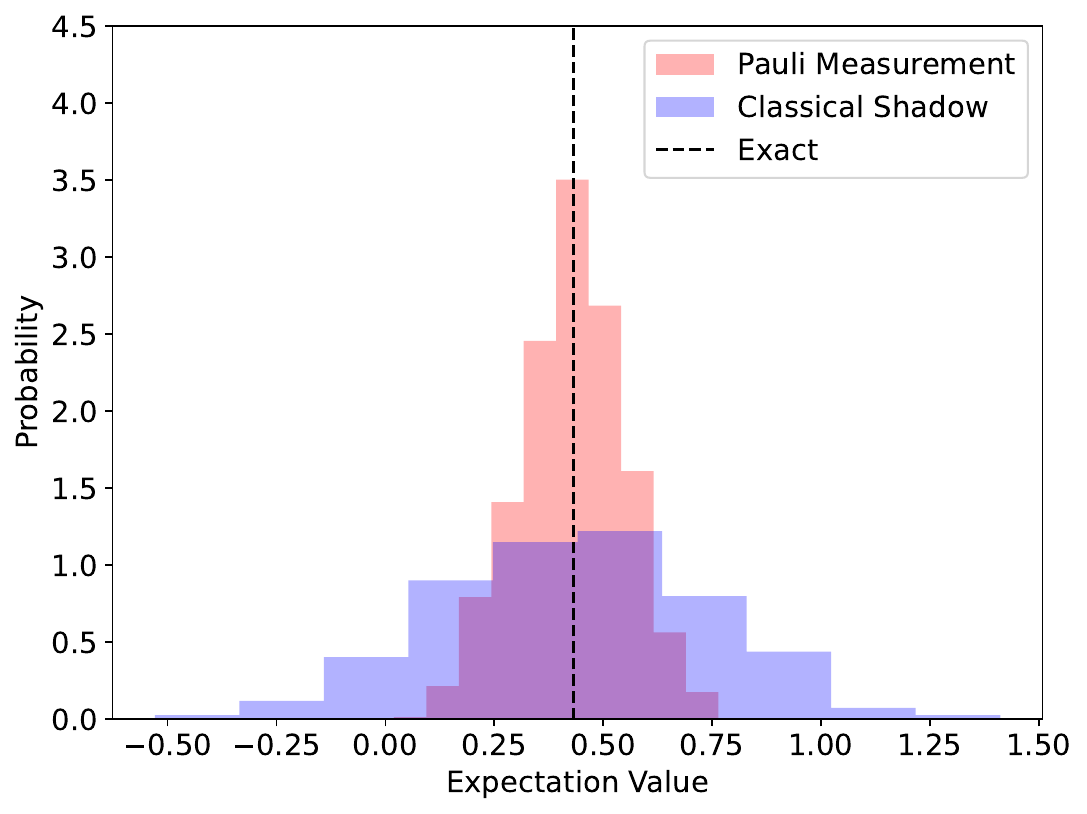}
    \caption{Comparison of expectation values obtained by (ordinary) Pauli-basis measurement and classical shadows. Here, expectation values of Pauli $Z$ on the first qubit is estimated 1000 times, where measurement shots for ordinary case is $N_{shot}=51$ and the number of snapshot for classical shadows is $N_{S}=51$ with $K=3$. 
    We use four-qubit RSP circuits with random parameters $\bm{\theta}$ to generate a quantum state.
    The variances are 0.016 and 0.072 for ordinary case and classical shadows, respectively.
    The dashed line express the exact expectation.}
    \label{fig:exp_comp}
\end{figure}

First, we consider the case where the molecule is at the equilibrium nuclear distance ($R=0.735$~\AA) and the parameters $\bm{\theta}^{*}$ are nearly optimal values which we obtain by performing the VQE optimization with BFGS.
This is actually the situation where the assumptions in the derivation of QCPMD hold.
Here, we run the simulation for 4,000 fs.
Fig.~\ref{fig:res1} illustrates the histograms of bond lengths obtained by the original QCPMD and the QCPMD with classical shadows.
Here, we get rid of the first 250 fs and use the rest of them to make the figures.
It turned out that both of them can reproduce the equilibrium state.
To evaluate the performance quantitatively, we also estimate the average of the bond length.
As a result, the bond lengths for QCPMD with and without classical shadows are 0.735~\AA~and 0.736~\AA, respectively, which are (almost) the same with the equilibrium distance, i.e., $0.735$~\AA.
The slight difference could be reduced if we run the simulation longer.
We note that the variance of bond length for classical shadow case is larger, because the uncertainty is increased by the randomness in choosing the measurement basis.
This can be easily checked by estimating expectation values using ordinary Pauli-basis measurement and classical shadows.
As a toy task, we here consider four-qubit RSP circuit with random parameters $\bm{\theta}$ and a Pauli $Z$ operators on the first qubit as the target observable.
Fig.~\ref{fig:exp_comp} clearly shows larger variance for classical shadows with $N_{S}=51$ and $K=3$ than measurement with $N_{shot}=51$; the variances averaged over five trials using different parameters $\bm{\theta}$ are 0.016 and 0.077 for ordinary Paili-basis measurement and classical shadows, respectively.

\begin{figure}[t]
    \centering
    \includegraphics[scale=0.48]{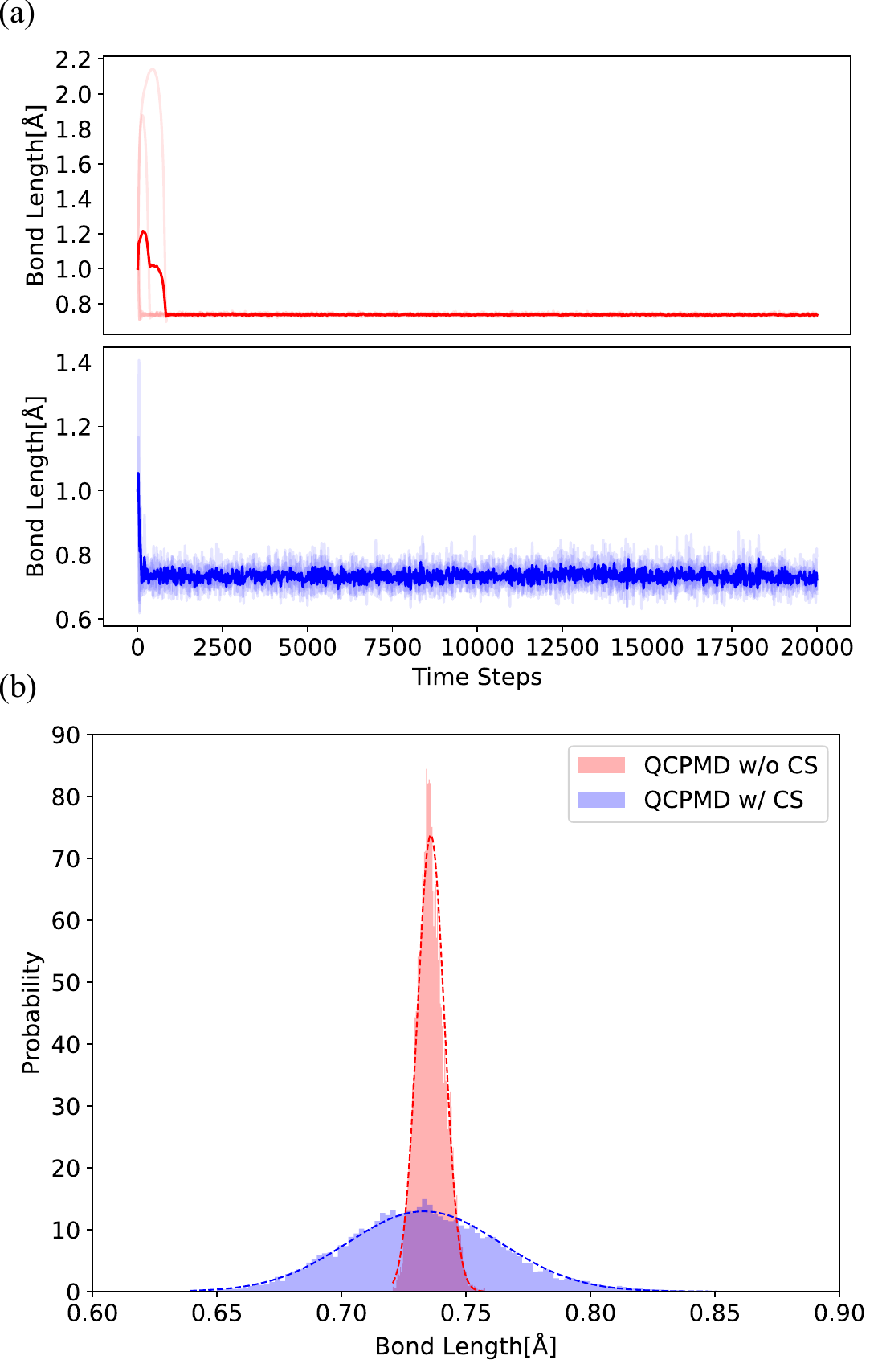}
    \caption{Simulation results produced by the original QCPMD (QCPMD w/o CS) and QCPMD with classical shadows (QCPMD w/ CS) in case the initial condition is far from the equilibrium state. (a) Trajectories of bond lengths for QCPMD without classical shadow (top) and QCPMD with classical shadow (bottom) are shown. The solid line represents the average trajectory for five trials with different initial parameters $\bm{\theta}$, while a transparent line illustrates the one for each trial. (b) Distributions of bond lengths obtained by QCPMD w/o CS and QCPMD w/ CS are demonstrated. Note that this is a result for a trial out of five.}
    \label{fig:res2}
\end{figure}

Secondly, we examine the case with $R=1.0$ and the parameters $\bm{\theta}$ are chosen at random.
We consider this situation to see the practical applicability of QCPMD.
That is, although the formula is derived based on the equilibrium state assumptions, we do not always know the optimal nuclei positions and electronic wavefunctions at the beginning.
Thus, we check the performance of QCPMD when we start the simulation with the random guess.
Here, the number of measurement shots and the number of snapshots are the same as in the first case.
We also prepare five different parameter sets for initial values of $\bm{\theta}$.
The total simulation time is set as 2,000 fs.
In Fig.~\ref{fig:res2}, we show the trajectory of bond length over the simulation time and its histogram after reaching equilibrium (bond lengths after 250 fs).
We observe that the QCPMD method can reach the equilibrium state after certain timesteps, irrespective of whether classical shadows are used or not.
We also find that the average of the bond lengths are 0.733~\AA~and 0.737~\AA~for the case with and without classical shadows, respectively.
These results suggest that QCPMD could perform simulations even when the initial condition is far away from the equilibrium.
Moreover, we could say classical shadows do not deteriorate the performance of QCPMD while realizing the resource efficiency.

Lastly, we discuss a practical difficulty of QCPMD.
As mentioned above, an advantage of the method is to use the statistical noise as the thermostat.
However, more samples would be required to realize the control.
Thermal control is actually reflected in QCPMD by tuning dissipation coefficients $\bm{\gamma}$ and $\bm{\zeta}$, which rely not only on the temperature, but also on the variance of forces.
This implies that the precise computation of the variance for the equilibrium state is necessary.
To this end, we should know the exact equilibrium state and have to compute the variance at that point.
These factors could hinder the practicality of QCPMD, because the fluctuation as shown in Figs.~\ref{fig:res1} and~\ref{fig:res2} might make it difficult to find the exact equilibrium state, and additional estimation for the variance is resource demanding.  
In our experiments, we avoid this by using fixed dissipation coefficients.
On the other hand, this implies that our results shown above might not reproduce the simulation at 70 K, but at certain ``effective" temperature from this viewpoint; the simulation by QCPMD with classical shadows results in equilibrium states at higher effective temperature than the original QCPMD, while the results by the original QCPMD might not be the ones at 70 K as well.
This fact also indicates that we also need to estimate the variance at each step in our second case, which could undermine the resource efficiency.
A possible approach to mitigating this challenge is to save forces at every step and then update the dissipation coefficients at some interval by computing the variance with the saved ones.
This might work under the assumption that quantum states do not change significantly during certain period. 
In addition, the classical shadow technique could be used for estimating the variance.
Thus, our scheme might be more advantageous in case we control the temperature exactly, which we leave for future work.

\section{Conclusion} \label{sec:conc}
In this work, we introduce the classical shadow technique to improve the resource efficiency of the QCPMD method for AIMD simulations.
We focus on the $\text{H}_2$ molecule and numerically demonstrate that the QCPMD with classical shadows can successfully reproduce the equilibrium state at a constant temperature.
This study will encourage the exploration of QCPMD' performance for simulating large systems and lead to a new invention of efficient AIMD simulation methods using NISQ devices. 
In addition, as we discussed in Section~\ref{sec:method}, some approaches such as grouping~\cite{yen2023deterministic} and derandomized classical shadows~\cite{huang2021efficient} could be more effective in this setting.
Thus, comparing these approaches would be intriguing.
Furthermore, it would be interesting to investigate the performance of our QCPMD scheme for larger systems.


\bibliographystyle{naturemag}
\bibliography{apssamp}

\end{document}